\documentclass{aa}

\usepackage{graphics}

\begin{document}

\thesaurus{01     %
           (11.19.3)  
           (13.07.1)} 


\title{Star-Forming Regions near GRB 990123\thanks{Based on
observations with the NASA/ESA \emph{Hubble Space Telescope}, obtained
at the Space Telescope Science Institute, which is operated by the
Association of Universities for Research in Astronomy, Inc., under
NASA contract NAS5-26555.}}

\author{Stephen Holland\inst{1}
           \and
        Jens Hjorth\inst{2}
       }

\offprints{Stephen Holland}

\institute{Institut for Fysik og Astronomi (IFA),
           Aarhus Universitet,
           Ny Munkegade, Bygning 520,
           DK--8000 {\AA}rhus C,
           Denmark \\
           e-mail: holland@obs.aau.dk
         \and
	   Astronomical Observatory,
           University of Copenhagen,
           Juliane Maries Vej 30,
           DK-2100 Copenhagen {\O},
           Denmark \\
           e-mail: jens@astro.ku.dk
          }

\date{Received; accepted}

\maketitle


\begin{abstract}

	We reduced the \emph{Hubble Space Telescope} Space Telescope
Imaging Spectrograph images of the gamma-ray burst \object{GRB 990123}
that were obtained on 8--9 February 1999 and find $V_0 = 25.36 \pm
0.10$, which corresponds to a flux of $0.258 \pm 0.023$ $\mu$Jy for
the optical transient 16.644 days after the burst's peak.  The
probable host galaxy has $V_0 = 24.25 \pm 0.07$ ($0.716 \pm 0.046$
$\mu$Jy) and the optical transient is located $0\farcs65$ ($= 5.5$
kpc) south of the galaxy's nucleus.  We fit and subtracted a scaled
point-spread function to the optical transient and found evidence for
three bright knots situated within $0\farcs5$ ($= 4.3$ kpc) of the
optical transient.  Each knot has $V_0 \sim 28.1 \pm 0.3$, a
rest-frame $V$-band luminosity of $L_V \sim 5$--$8 \times 10^8
L_{\sun}$, and a star-formation rate of at least $\sim 0.1$--$0.2
{\cal M}_{\sun}$ yr$^{-1}$.  The knots are centrally concentrated with
full-width at half-maximum of $\sim 0\farcs17$ ($= 1.5$ kpc).  Their
sizes and luminosities are consistent with their being star-forming
regions.  The optical transient is located $0\farcs15$ ($= 1.3$ kpc)
southeast of the centre of one of these knots.


\end{abstract}


\section{Introduction\label{intro}}

	On 23 January 1999 \emph{BeppoSAX} detected \object{GRB
990123}, one of the strongest recorded gamma-ray bursts (GRBs) (Heise
\cite{H99}; Piro \cite{P99}).  The optical transient (OT) associated
with this GRB reached a peak magnitude of $V = 8.95$ (Akerlof \& McKay
\cite{AM99}) only 47 seconds after the initial detection.
Spectroscopic observations with the Nordic Optical Telescope (Hjorth
et~al.\ \cite{HA99}; Andersen et~al.\ \cite{AC99}), and the Keck
Telescope (Kelson et~al. \cite{KI99}; Kulkarni et~al.\ \cite{KD99}),
showed strong metallic absorption lines with redshifts of $z=1.600$,
suggesting that the OT is located either in, or behind, an absorbing
system at $z = 1.6$.  Andersen et~al.\ (\cite{AC99}) placed an upper
limit of on the redshift of $z < 2.05$ based on ultraviolet photometry
and the absence of Lyman-$\alpha$ forest lines.  For a Friedman
cosmology with $H_0 = 65$ km s$^{-1}$, $\Omega_0 = 0.2$, and $\Lambda
= 0$ a redshift of $z = 1.6$ corresponds to $11\,982$ Mpc and a
distance modulus of $\mu = 45.39$.  This means that the OT reached a
peak absolute $V$-band magnitude of $M_V \sim -36.5$\footnote{All
magnitudes are in the rest frame of the observer, unless otherwise
noted.}, which gives an intrinsic luminosity of $L_V \sim 2 \times
10^{16} L_{\sun}$, approximately $10^7$ times brighter than a Type I
supernova.  For an isotropic burst the implied energy release
approached $3$--$4.5 \times 10^{54}$ erg, using BATSE fluence (Kippen
\cite{K99}).

	There are two broad families of models for GRBs, both of which
are related to the end stages of the lives of massive stars, and both
of which predict that GRBs should trace the star-formation rate in the
Universe (e.g.\ Wijers et~al.\ \cite{WB98}). The first family is the
exploding object family, which includes the ``failed supernova'' model
of Woosley (\cite{W93}) and the ``hypernova'' model of Paczy{\'n}ski
(\cite{P98}).  These models predict that the progenitors of GRBs are
short-lived objects with low space velocities, so GRBs should be found
within $\sim 0.5$ kpc of the star-forming regions where the
progenitors formed.  The second family of models is the binary
progenitor family, which includes the binary neutron star (NS-NS)
model of Narayan et~al. (\cite{NP92}) and the merger of a black hole
and a neutron star (BS-NS) model of Paczy{\'n}ski (\cite{P91}).  These
models predict that GRB progenitors will have high space velocities,
due to two supernova explosions having occurred in the progenitor's
binary system, so the GRB could be found a significant distance from
the star-forming region where the progenitor was formed.  OTs have
been associated with several GRBs, and most of these OTs have been
located within $1\arcsec$ of a faint galaxy.  Redshifts have been
measured for four of these candidate host galaxies and all are located
at cosmological distances.


\section{The Data\label{SECTION:data}}

	\emph{Hubble Space Telescope} (\emph{HST}) observations of the
OT associated with \object{GRB 990123} were made between 23:06:54 UT
on 8 February 1999 and 03:21:43 UT on 9 February 1999 as part of the
Cycle 8 proposal GO-8394 (Beckwith \cite{B99}) in order to identify
the host galaxy and to study the region around the OT\@.  These
observations consisted of six 1300 second exposures taken with the
Space Telescope Imaging Spectrograph (STIS) in its Clear Aperture
(50CCD) mode.  Each exposure was split into two 650 second
sub-exposures to allow for the removal of cosmic rays.  A six-position
spiral dithering pattern, with offsets of $\sim 10$ pixels ($\sim
0\farcs5$), was used.  The CCD gain was set to 1 $\mathrm{e}^-$/ADU,
and the read-out noise was 4 $\mathrm{e}^-$/pixel.  The data was
processed through the standard STIS pipeline and immediately made
available to the astronomical community.

	Fruchter et~al.\ (\cite{FS99}) reported a preliminary
magnitude of $V = 25.4 \pm 0.1$ for the OT based on the \emph{HST}
images.  Bloom et~al.\ (\cite{BO99}) used this data to derive
magnitudes for the OT and for the purported host galaxy, and noted
that there may be a small star-forming region to the north of the
OT\@.  Fruchter et~al.\ (\cite{FT99}) also derived magnitudes for the
OT and the host galaxy, and determined a decay rate for the flux from
the OT which suggests that the light from the OT may have been beamed
for the first $\sim 2$ days after the burst, and then made a
transition to and isotropic expansion.

\begin{figure}
\resizebox{\hsize}{!}{\includegraphics{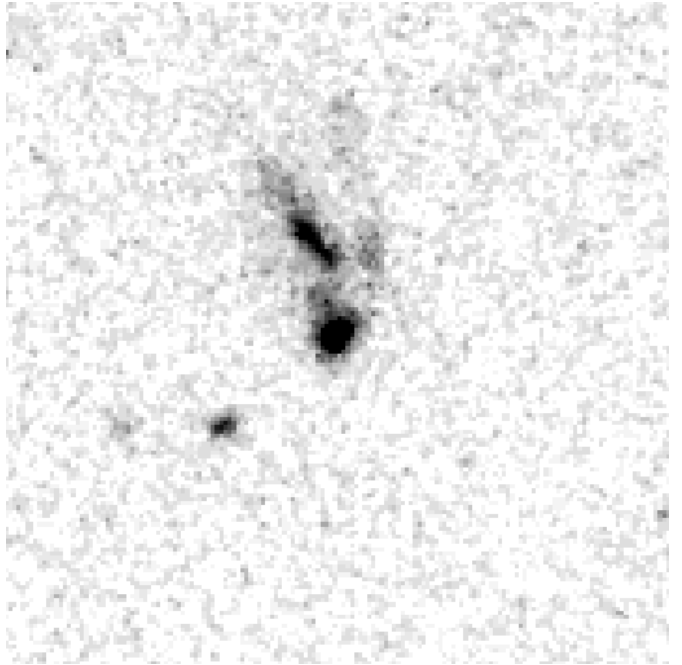}}
\caption{This figure shows the central $3\farcs5 \times 3\farcs5$
section of the drizzled image.  The scale is $0\farcs0254$/pixel,
north is towards the top of the image, and east is towards the left.
The OT is the bright object in the centre of the image.  The nucleus
of the probable host galaxy is the centroid of the elongated object
that lies to the north of the OT and extends towards north-northeast.}
\label{FIGURE:host_ot}
\end{figure}

	We combined the six STIS images using the {\sc Dither} (v1.2)
software (Fruchter \& Hook~\cite{FH99}) as implemented in
IRAF\footnote{Image Reduction and Analysis Facility (IRAF), a software
system distributed by the National Optical Astronomy Observatories
(NOAO).} (v2.11.1)/STSDAS (v2.0.2).  A ``pixfrac'' value of 0.5 was
used, and the final output image had a scale of $0\farcs0254$/pixel.
Fig.~\ref{FIGURE:host_ot} shows the details of the OT and the probable
host galaxy. We found the location of the OT to be $\alpha =
15^\mathrm{h} 25^\mathrm{m} 30\fs30, \delta = +44\degr 45\arcmin
59\farcs1$ (J2000 coordinates) based on the World Coordinate System
information in the STIS image headers.  We have taken the centroid of
the bright elliptical object to the north of the OT (Bloom et~al.\
[\cite{BO99}]'s object A) to be the nucleus of the host galaxy.  It is
located at $\alpha = 15^\mathrm{h} 25^\mathrm{m} 30\fs32, \delta
+44\degr 45\arcmin 59\farcs7$.  The OT is situated $0\farcs65$ ($=
5.5$ kpc) south of the nucleus of the host galaxy ($174\fdg48$ east of
north).

\section{The Magnitude of the Optical Transient\label{SECTION:magnitudes}}

	We estimated the total $AB$ magnitude of the OT by performing
aperture photometry with an aperture radius of $0\farcs28$.  The local
sky was measured in an annulus with an inner radius of $0\farcs28$ and
a width of $0\farcs08$.  The photometry aperture and sky annulus were
chosen to maximize the signal-to-noise from the OT and to minimize the
contamination from the host galaxy and the nebulosity seen near the
OT\@.  The integrated magnitude depends slightly ($\pm \sim 0.1$ mag)
on the sizes of the aperture and sky annulus since the OT is situated
near several bright knots in the outer regions of the host galaxy.  No
aperture correction was applied since the uncertainties in the
background subtraction are larger than the aperture correction.  The
STIS Instrument Handbook gives a zero point in the $AB$ system of
$26.386$ for the 50CCD (clear) imaging mode, which corresponds to a
zero-point flux of $F_{\nu,0} = 0.1013$ $\mu$Jy.  This zero point
gives a calibrated $AB$ magnitude for the OT of $25.37 \pm 0.10$,
which corresponds to a flux of $0.259 \pm 0.024$ $\mu$Jy. We assume
that the OT has a power-law spectrum of the form $F_{\nu,0} = k
\nu^{\beta}$ where $\beta = -0.8$ (Bloom et~al.\ \cite{BK99}) and the
zero-point flux for the STIS in its 50CCD (clear) imaging mode yields
$k = 5.936 \times 10^{10}$ $\mu$Jy.  Fruchter et~al.\ (\cite{FT99})
note that the STIS magnitude of an object with a power-law spectrum is
largely independent of $\beta$ due to the symmetric shape of the 50CCD
(clear) bandpass.  We converted the $AB$ magnitude to the Johnson
$V$-band and Kron-Cousins $R$-band using this power-law spectrum and
the photometric zero-points from Fukugita et~al.\ (\cite{FS95}).  The
resulting calibration equations are $V = 26.429 - 2.5\log_{10}(C)$ and
$R = 26.083 - 2.5\log_{10}(C)$, where $C$ is the observed count rate
in ADU/s on the drizzled image.

	The OT is at $b^\mathrm{II} = +54\fdg64$, $l^\mathrm{II} =
73\fdg12$.  Schlegel et~al.\ (\cite{SF98}) find the Galactic reddening
in this direction to be $E_{B\!-\!V} = 0.016$.  We used $R_V = 3.09
\pm 0.03$ (Rieke \& Lebofsky \cite{RL85}) and $A_R/A_V = 0.8686 -
0.3660/R_V$ (Cardelli et~al.\ \cite{CC89}) to derive extinction
corrected magnitudes of $V_0 = 25.36 \pm 0.10$, and $R_0 = 25.03 \pm
0.10$ for the OT\@.  These correspond to fluxes of $0.258 \pm 0.023$
$\mu$Jy and $0.294 \pm 0.027$ $\mu$Jy respectively.


\section{Star-Forming Regions Near GRB 990123\label{SECTION:star_form}}

\begin{figure}
\resizebox{\hsize}{!}{\includegraphics{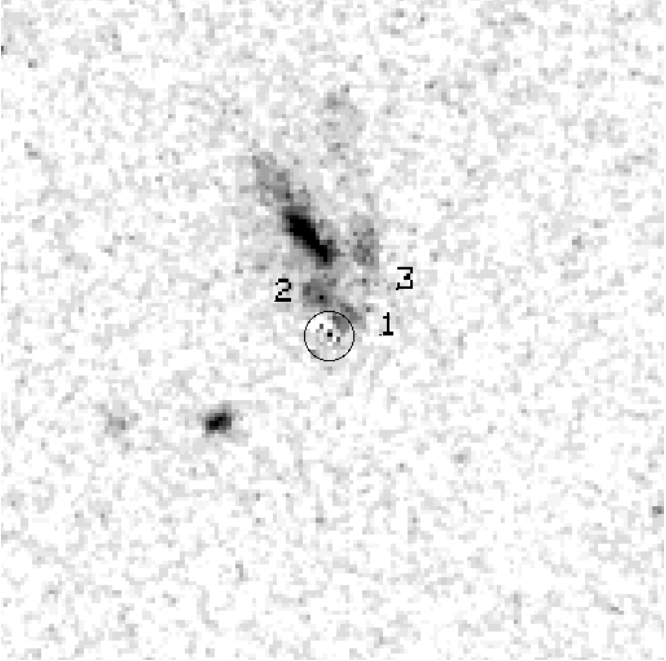}}
\caption{This is the same image as Fig.~\ref{FIGURE:host_ot} except
the OT has been subtracted.  The location of the OT is indicated by
the circle.  Three knots are visible to the north and northwest of the
OT (see Table~\ref{TABLE:knots}).  Knots \#2 and \#3 were identified
as A2 and A1 respectively by Bloom et~al.\ (\protect\cite{BO99}).  The
scale is $0\farcs0254$/pixel, north is to the top of the image, and
east is towards the left.}
\label{FIGURE:host}
\end{figure}

	In order to examine the substructure near the OT we used {\sc
DaoPhot II} (Stetson \cite{S87}) to subtract the light from the OT\@.
We constructed a point-spread function (PSF) using STIS CCD
observations of a star field in the Galactic globular star cluster
\object{Omega Centaurus}.  These images were taken with the STIS in
the 50CCD (clear) imaging mode as part of the Cycle 7 proposal 7079
and were combined and drizzled in exactly the same way that the images
of \object{GRB 990123} were (output scale $= 0\farcs0254$/pixel,
``pixfrac'' $= 0.5$).  Fig.~\ref{FIGURE:host} shows the STIS image of
the region around the OT with the light from the OT subtracted.

	The host galaxy appears to be a large irregular galaxy with
several non-axisymmetric components.  The total $AB$ magnitude of the
galaxy (excluding the OT) is $24.31 \pm 0.07$.  Castro-Tirado et~al.\
(\cite{CT99}) find a flat spectrum ($\beta \sim 0$) consistent with a
starburst galaxy.  Adopting this spectrum, and assuming no internal
reddening in the galaxy, we find $V_0 = 24.25 \pm 0.07$ and $R_0 =
24.07 \pm 0.07$ for the host galaxy, corresponding to fluxes of $0.716
\pm 0.046$ $\mu$Jy and $0.711 \pm 0.046$ $\mu$Jy respectively.  The
total luminosity is $L_V = (2.4 \pm 0.2) \times 10^{10} L_{\sun} =
(1.0 \pm 0.1) L^{\ast}_V$.  There is a large amount of unresolved
light around the galaxy, and an arc-like feature can be seen $\sim
0\farcs4$ ($= 3.5$ kpc) west of the nucleus.  This structure is $\sim
1\arcsec$ ($= 8.6$ kpc) long and $\sim 0\farcs2$ ($= 1.7$ kpc) wide
with a bright knot (\#3 $=$ A1 [Bloom et~al.\ \cite{BO99}]) at its
southern end.  Two more small knots can be seen to the north (\#2 $=$
A2 [Bloom et~al.\ \cite{BO99}]) and northwest of the OT (\#1).  The
locations, and distances from the OT, for these knots are listed in
Table~\ref{TABLE:knots}.  The final column gives the Bloom et~al.\
(\cite{BO99}) identifications.  The magnitudes, fluxes, and rest-frame
$V$-band luminosities for each knot are listed in
Table~\ref{TABLE:knot_data}.  The uncertainties in the magnitudes and
fluxes are $\sim 0.3$ mag and $\sim 5.6$ nJy respectively.  The
uncertainties in the total luminosities are $\sim 2 \times 10^8
L_{\sun}$.

\begin{table}
\begin{center}
\caption{The locations of possible star-forming regions near the OT.}
\smallskip
\begin{tabular}{cccccc}
\hline
\hline
\vspace{0.1 em}
Knot & $\alpha_\mathrm{J2000}$ & $\delta_\mathrm{J2000}$ & $d$ & $d$ (kpc) & ID\\
\hline
\vspace{0.1 em}
 1 & 15:25:30.30 & $+$44:45:59.2 & $0\farcs15$ & 1.3 & $\cdots$ \\
 2 & 15:25:30.31 & $+$44:45:59.3 & $0\farcs25$ & 2.2 & A2 \\
 3 & 15:25:30.28 & $+$44:45:59.5 & $0\farcs48$ & 4.1 & A1 \\
\hline
\hline
\end{tabular}
\label{TABLE:knots}
\end{center}
\end{table}

\begin{table}
\begin{center}
\caption{Properties of the possible star-forming regions near the OT.}
\smallskip
\begin{tabular}{cccccccc}
\hline
\hline
\vspace{0.1 em}
Knot & $V_0$ & nJy & $R_0$ & nJy & $L_V/L_{\sun}$ & SFR \\
\hline
\vspace{0.1 em}
 1 & 28.3 & 17.2 & 28.1 & 17.4 & $5.7 \times 10^8$ & 0.09 \\
 2 & 28.1 & 20.7 & 27.9 & 20.9 & $7.0 \times 10^8$ & 0.11 \\
 3 & 28.0 & 22.7 & 27.8 & 22.9 & $7.7 \times 10^8$ & 0.12 \\
\hline
\hline
\end{tabular}
\label{TABLE:knot_data}
\end{center}
\end{table}

	The central regions of all three knots are well fit by the
PSF, which suggests that the knots are centrally concentrated.  The
drizzled image has a resolution of $0\farcs0254$/pixel ($= 0.2$
kpc/pixel).  Therefore, we conclude that most of the light is
concentrated in the inner $0.2$ kpc of each knot.  This is comparable
with the sizes of star-forming regions and H$\;${\small II} regions in
the local Universe.  The PSF does not, however, provide a good fit to
the outer regions of the knots.  There is excess light left in the
images of each knot after the scaled PSFs have been subtracted.  This
suggests that the knots consist of dense, centrally concentrated cores
embedded in extended structures.  The apparent full-width at
half-maximum (FWHM) of each knot is $\sim 0\farcs18$ while the PSF has
FWHM $= 0\farcs036$.  Correcting the apparent knot diameters for the
width of the PSF gives intrinsic FWHMs of $\sim 0\farcs17$ ($\sim 1.5$
kpc) for the knots.  This is consistent with the knots containing
embedded star-forming regions.

	Knot \#1 is partially covered by the OT and not clearly
visible until the OT is subtracted.  We believe that this knot is a
real feature and not an artifact of the PSF subtraction since similar
features are not seen when the PSF is used to subtract stars from the
\object{Omega Centaurus} images.  Fig.~\ref{FIGURE:host} shows no
systematic change in intensity between the northeast half of the knot,
which was not obscured by the OT, and the southwest half, which was
obscured by the OT\@.  This also suggests that the knot is not an
artifact of the PSF subtraction.  To test the ability of PSF
subtraction to reveal structure under the OT we generated a series of
artificial stars with the same magnitude as the OT\@.  These stars
were put on the three knots, and in empty parts of the image near the
OT, then PSFs were fit and subtracted for each artificial star.  We
found that the knots were clearly visible after the artificial stars
were removed, although the knots located under artificial stars
appeared less centrally concentrated after the artificial stars were
subtracted.  Comparing the recovered magnitudes of the isolated
artificial stars with those of the artificial stars situated on the
knots suggests that we are able to detect knots with $V_0 < 29$ that
are located under the OT\@.

	The OT for \object{GRB 990123} is located on the southeast
edge of knot \#1.  This knot is the most likely source of the metallic
absorption lines seen in the spectra of the OT (Andersen et~al.\
\cite{AC99}; Kulkarni et~al.\ \cite{KD99}), although we can not rule
out the possibility that the absorption lines are due to a very small,
undetected, faint knot located under the OT\@.  Such absorption
systems are often associated with high column densities of hydrogen,
which in turn are associated with star formation.  We used Eq.~2 of
Madau et~al.\ (\cite{MP98}) to estimate the star-formation rate (SFR)
in each of the knots listed in Table~\ref{TABLE:knots}, assuming a
flat ($\beta = 0$) spectrum.  For $z = 1.6$ a rest-frame wavelength of
$\lambda_0 = 2800$ {\AA} corresponds to an observed wavelength of
$\lambda = 7280$ {\AA}, so we computed the observed flux at $\lambda =
7280$ {\AA} by extrapolating between the $V$- and $R$-band fluxes.
These were computed for each knot in the same manner as was done for
the host galaxy (see Sect.~\ref{SECTION:magnitudes}).  The estimated
SFRs for each knot, assuming a Salpeter initial mass function, are
listed in Table~\ref{TABLE:knot_data}.  For a Scalo initial mass
function multiply the SFR by 1.55.  The uncertainty in each SFR is
$\sim 0.03$ ${\cal M}_{\sun}$ yr$^{-1}$.  These SFRs assume that there
is no dust, or obscured star formation, in the knots.  This is
probably a poor assumption if the knots are star-forming regions.
Therefore, our derived SFRs should be considered a lower limit on the
true SFR in each knot.

	The OT is located at a projected distance of $0\farcs65$ ($=
5.6$ kpc) from the nucleus of the host galaxy.  Bloom et~al.\
(\cite{BS98}) calculated that $\sim 50$\% of the GRBs from NS-NS and
BS-NS progenitors in galaxies with a shallow gravitational potential
will occur within 5 kpc of the nucleus of the host galaxy and $\sim
90$\% will occur within 30 kpc.  This is consistent with the location
of the OT relative to the nucleus of the host galaxy.  If the
progenitor is a failed supernova, or a hypernova, then the OT should
be located within a few hundred parsecs of the star-forming region
(Paczy{\'n}ski \cite{P98}).  The \object{GRB 990123} OT is located at
a projected distance of $0\farcs15$ ($= 1.3$ kpc) southeast of the
centre of the nearest knot (\#1), which is larger than the expected
separation if the GRB was due to the explosion of a massive star, yet
consistent with the NS-NS and BS-NS hypotheses.  We wish to stress
that this conclusion depends on there being no faint star-forming
region directly under the OT\@.  Further observations will be needed,
after the OT has faded, to determine if there are other small star
formation regions that are currently hidden by the OT\@.


\begin{acknowledgements}

We would like to thank S. Beckwith of the Space Telescope Science
Institute for making the \emph{HST}/STIS observations of \object{GRB
990123} available to the astronomical community.  We would also like
to thank H. Ferguson for his useful comments.  S. H. is supported by
the Danish Centre for Astrophysics with the \emph{HST}.  This work was
supported by the Danish Natural Science Research Council.
\end{acknowledgements}


\end{document}